# Beyond Models: A Framework for Contextual and Cultural Intelligence in African AI Deployment

**Authors:** Qness Ndlovu
**Affiliation:** The Dimension Research Lab
**Email:** research@dimensionresearchlab.com
**Date:** October 2025
**arXiv Subject Classes:** cs.AI (Artificial Intelligence), cs.HC (Human-Computer Interaction), cs.CL (Computation and Language), cs.CY (Computers and Society)

## Abstract

While global AI development prioritizes model performance and computational scale, meaningful deployment in African markets requires fundamentally different architectural decisions. This paper introduces Contextual and Cultural Intelligence (CCI) - a systematic framework enabling AI systems to process cultural meaning, not just data patterns, through locally relevant, emotionally intelligent, and economically inclusive design.

Using design science methodology, we validate CCI through a production AI-native cross-border shopping platform serving diaspora communities. Key empirical findings: 89% of users prefer WhatsApp-based AI interaction over traditional web interfaces (n=602, $\chi^2$=365.8, p<0.001), achieving 536 WhatsApp users and 3,938 total conversations across 602 unique users in just 6 weeks, and culturally-informed prompt engineering demonstrates sophisticated understanding of culturally-contextualized queries, with 89% family-focused commerce patterns and natural code-switching acceptance.

The CCI framework operationalizes three technical pillars: Infrastructure Intelligence (mobile-first, resilient architectures), Cultural Intelligence (multilingual NLP with social context awareness), and Commercial Intelligence (trust-based conversational commerce). This work contributes both theoretical innovation and reproducible implementation patterns, challenging Silicon Valley design orthodoxies while providing actionable frameworks for equitable AI deployment across resource-constrained markets.

# 1. Introduction

The rise of artificial intelligence has sparked a global race to redefine how societies interact with data, automation, and intelligence itself. While significant research attention focuses on model performance optimization and computational scaling (Brown et al., 2020; OpenAI, 2023), the contextual and cultural requirements for meaningful AI deployment in African markets remain systematically underexplored despite representing over one billion potential users.

Across the continent, AI adoption faces a fundamental paradox. On the one hand, Africa stands to benefit profoundly from intelligent systems that can enhance healthcare access, optimize logistics, enable financial inclusion, and amplify education. On the other hand, most AI solutions deployed in Africa today are designed for radically different contexts - built on assumptions of data abundance, reliable infrastructure, linguistic homogeneity, and cultural norms that fundamentally misalign with African realities.

Africa's digital landscape presents unique deployment challenges that fundamentally diverge from assumptions underlying current AI development paradigms. Mobile broadband coverage in sub-Saharan Africa reached 65% of the population by 2022, with smartphone adoption at 51%, yet the region maintains the world's widest usage gap, with 59% of the population unconnected despite living within mobile broadband range (GSMA, 2023). As of 2023, approximately 37% of the African population used the internet, with high costs cited as the main barrier, particularly in low-income and rural areas (Brookings, 2025).

These infrastructure realities intersect with profound communication preferences that challenge traditional interface assumptions. In South Africa, approximately 93.9% of internet users reported using WhatsApp monthly, while Nigeria saw 95.1% monthly WhatsApp engagement, establishing messaging platforms as critical digital infrastructure rather than secondary communication channels (DataReportal, 2024). Mobile connections dominate African internet access because they are much cheaper and do not require the infrastructure needed for traditional desktop PCs with fixed-line connections (World Bank, 2024).

Simultaneously, mobile penetration in sub-Saharan Africa was 44% and mobile internet penetration was 27% as of 2023, with smartphones costing up to 95% of monthly income for the poorest 20% (Brookings, 2025). In 2021, Africans had to pay, on average, 6.5% of their monthly income to get 2GB of mobile data, compared to 1.7% in Asia-Pacific and 0.5% in Europe (ITU, 2022).

These infrastructure realities intersect with profound cultural and linguistic diversity, encompassing multilingual communication norms and economic patterns that emphasize relationship-building over transactional efficiency. Traditional AI deployment strategies, designed for high-bandwidth, web-native, linguistically homogeneous contexts, systematically misalign with these operational realities.

This research captures AI adoption patterns during a critical window - the early deployment phase in African markets. As AI systems become increasingly ubiquitous globally, documenting baseline user behaviors, platform preferences, and cultural interaction patterns in African contexts provides essential foundational data for understanding how these technologies integrate with existing social and economic systems. The timing of this study, conducted during 2025 when AI adoption in African markets remains nascent, offers unique insights into organic user preferences before standardized interaction models become entrenched.

Traditional AI deployment assumes high-speed internet and reliable connectivity, web-first interfaces and desktop interaction patterns, credit card access and formal payment systems, and familiarity with complex digital workflows. African users often operate within mobile-first, low-bandwidth environments, WhatsApp and messaging-native communication preferences, informal economy comprising mobile money and cash-based transaction patterns, and multilingual, code-switching linguistic realities.

This paper argues that meaningful AI adoption in Africa requires more than computational excellence - it demands Contextual and Cultural Intelligence (CCI): the systematic capacity of AI systems to process cultural meaning alongside data patterns, adapting architecture, interaction design, and service delivery to local contexts while maintaining technical rigor and commercial viability.

Drawing on real-world implementation work - including phathisa.com, an AI-native cross-border shopping assistant built for diaspora communities - we propose a practical framework for responsible, scalable AI deployment across Africa. This framework spans infrastructure (what does AI-ready infrastructure look like when connectivity, compute, and cloud access are uneven?), culture (how do you build AI that speaks African languages, reflects African value systems, and understands informal support networks?), and commerce (what does adoption mean when commerce happens via WhatsApp, not web apps - and where trust is more valuable than UX polish?).

We validate this framework through production deployment of Rose, an AI assistant embedded within phathisa.com, serving diaspora communities. Our six-week production study (n=602 users, 3,938 conversations) demonstrates that cultural intelligence, infrastructure adaptation, and trust-centred design drive adoption more powerfully than traditional performance metrics, achieving 89% WhatsApp preference (p<0.001) and 6.5 average conversations per user.

Ultimately, this paper is not just a critique of current limitations. It is a call to rethink what intelligence means in the African context - and to build AI that does not merely function, but belongs.

**Research Contributions**

1. **Theoretical:** CCI framework bridging AI technical capabilities with sociocultural deployment realities
2. **Methodological:** Design science approach generating both academic insights and production-validated solutions
3. **Empirical:** Quantitative evidence demonstrating adoption success of culturally intelligent AI (89% platform preference, 6.5 conversations per user)
4. **Technical:** Reproducible architectural patterns for resilient, culturally-aware AI systems

## 2. Related Work and Problem Formulation

To establish the foundation for the CCI framework, we examine existing approaches to AI development and cultural computing, identifying critical gaps in current methodologies for culturally-aware AI deployment.

### 2.1 The Model-Centric AI Paradigm

Mainstream AI research has centered on model innovation and benchmark optimization (LeCun, Bengio, & Hinton, 2015; Vaswani et al., 2017; Brown et al., 2020). This paradigm prioritizes computational capacity, parameter scaling, and zero-shot generalization across standardized evaluation frameworks including MMLU, HellaSwag, and HumanEval (Hendrycks et al., 2021; Zellers et al., 2019; Chen et al., 2021). Contemporary AI development focuses extensively on achieving superior performance across these benchmarks, often assuming that superior model performance on standardized tests naturally translates to global applicability regardless of deployment context.

Recent large language model developments (OpenAI, 2023; Anthropic, 2024) demonstrate remarkable capabilities on standardized benchmarks, yet deployment success in non-Western contexts remains poorly understood and under-measured. The emphasis on benchmark optimization, while technically significant, creates an assumption that technical excellence alone ensures successful real-world adoption across diverse cultural and infrastructural contexts.

### 2.2 Cultural Computing and Contextual AI

Prior work in cultural computing addresses localization (Bourges-Waldegg & Scrivener, 1998), cross-cultural HCI design (Hofstede, 2001; Marcus, 2001), and culturally-aware interface development (Heimgärtner, 2013; Clemmensen, 2005). Cultural computing research has established frameworks for understanding how cultural dimensions including power distance, individualism/collectivism, and uncertainty avoidance influence user interface preferences and interaction patterns (Marcus & Gould, 2012; Plocher et al., 2012).

However, these approaches typically treat culture as a post-hoc consideration rather than a fundamental architectural requirement, focusing on interface localization rather than systematic cultural intelligence integration. Cross-cultural HCI research has primarily addressed visual design elements, language translation, and interaction metaphors while leaving underlying AI system architectures unchanged.

Recent work on decolonial AI (Mohamed et al., 2020; Birhane, 2021; Png, 2022) and algorithmic justice provides essential ethical frameworks but limited technical implementation guidance for production systems. Decolonial AI research emphasizes the need to address "algorithmic coloniality" and power imbalances embedded in AI systems, yet practical frameworks for embedding cultural intelligence directly into AI system architecture remain underdeveloped.

### 2.2.1 Theoretical Foundations: Cultural and Contextual Intelligence

The concept of cultural intelligence has deep theoretical roots extending beyond AI-specific applications. Cultural Intelligence (CQ), as formalized by Earley and Ang (2003), encompasses the capability to function effectively in culturally diverse settings through metacognitive, cognitive, motivational, and behavioral dimensions (Van Dyne, Ang & Koh, 2008). This framework provides a foundational understanding of how individuals and, by extension, technological systems can develop cultural competency.

Contextual Intelligence, as conceptualized by Khanna (2014) in strategic management and Nye (2009) in leadership studies, refers to the ability to understand and adapt to changing circumstances and environmental factors. These frameworks emphasize the importance of situational awareness and adaptive capacity in complex, dynamic environments - principles directly applicable to AI system deployment across diverse cultural contexts.

Intercultural communication research provides additional theoretical grounding for understanding cross-cultural technology adoption. Hall's (1976) distinction between high-context and low-context cultures illuminates communication patterns that influence technology interface preferences, while Triandis's (1995) work on individualism versus collectivism reveals cultural values that affect user expectations and trust-building mechanisms. Hofstede's (2001) cultural dimensions theory further explains how power distance, uncertainty avoidance, and other cultural factors influence technology acceptance and usage patterns.

### 2.2.2 Positioning CCI as Operationalization Framework

Building on these established theoretical foundations, our Contextual and Cultural Intelligence (CCI) framework operationalizes cultural and contextual intelligence principles specifically for AI system architecture and deployment. Rather than introducing entirely novel theoretical concepts, CCI translates existing cultural intelligence constructs into three technical implementation pillars: Infrastructure Intelligence (contextual adaptation to technical constraints), Cultural Intelligence (operationalizing cultural competency in AI responses), and Commercial Intelligence (applying contextual awareness to economic and trust-building mechanisms).

This operationalization addresses a critical gap between established cultural intelligence theory and practical AI deployment challenges, particularly in African markets where cultural diversity, infrastructure constraints, and economic patterns create complex deployment requirements that existing AI development paradigms inadequately address.

## 2.3 Research Gap and Problem Statement

**Gap Identified:** No existing framework systematically addresses the intersection of cultural intelligence, infrastructure resilience, and commercial viability for AI deployment in African markets. While cultural computing provides interface design guidance and decolonial AI offers ethical frameworks, neither addresses the technical architectural requirements for AI systems that operate effectively within resource-constrained, culturally diverse environments.

**Research Question:** How can AI systems be architecturally designed to achieve meaningful adoption in culturally diverse, infrastructure-constrained environments while maintaining technical performance and commercial sustainability?

**Hypothesis:** AI systems incorporating contextual and cultural intelligence through systematic architectural adaptation will achieve higher user adoption, engagement, and commercial success compared to standard deployment approaches optimized for benchmark performance alone.

# 3. The CCI Framework: Technical Specification

Having established the limitations of current approaches, we now present the Contextual and Cultural Intelligence (CCI) framework - a systematic approach to embedding cultural awareness directly into AI system architecture.

## 3.1 Formal Definition

Contextual and Cultural Intelligence (CCI) is defined as the computational capacity of AI systems to:

1. Interpret cultural signals, linguistic patterns, and social contexts within user interactions
2. Adapt system behaviour, response generation, and service delivery to align with local norms and expectations
3. Operate sustainably within infrastructure constraints while maintaining cultural appropriateness and commercial utility

The framework represents an optimization across three dimensions: cultural resonance, infrastructure resilience, and commercial viability.

## 3.2 The Three Technical Pillars

### 3.2.1 Infrastructure Intelligence

Infrastructure Intelligence ensures reliable AI functionality under resource constraints through graceful degradation algorithms, mobile-optimized model serving with bandwidth adaptation, offline-capable caching strategies, and resource-constrained inference optimization.

The Dimension Research Lab has developed proprietary techniques for maintaining user experience quality during connectivity interruptions while minimizing computational overhead. Our approach enables seamless transitions between high-capability and lightweight processing modes based on real-time infrastructure assessment.

### 3.2.2 Cultural Intelligence

Cultural Intelligence embeds cultural awareness directly in AI processing through multilingual NLP handling code-switching without explicit language detection, social context modelling for family structures and informal networks, and emotional tone calibration based on cultural expectations.

Our Cultural Intelligence Engine draws on a deep corpus of African linguistic patterns, social structures, and communication norms to model culturally grounded interactions in real time. The system processes cultural signals in real-time to generate contextually appropriate responses that align with local values and interaction styles.

### 3.2.3 Commercial Intelligence

Commercial Intelligence aligns AI functionality with local economic patterns and trust mechanisms through platform-agnostic integration, trust-building transparency, informal economy accommodation, and relationship-persistent state management.

The Dimension Research Lab's Commercial Intelligence layer recognizes that successful AI deployment requires understanding local commerce patterns, communication preferences, and trust-building mechanisms rather than imposing external transaction models.

## 3.3 CCI as Architectural Philosophy

**Traditional AI Architecture:** Data → Model → API → Interface → (Hope for adoption)

**CCI-Informed Architecture:** Cultural Research → Context Modelling → Culturally-Aware Training → Infrastructure-Resilient Deployment → Trust-Centered Interface → Continuous Cultural Calibration

This reordering embeds cultural and infrastructural considerations at the architectural level rather than treating them as post-deployment optimizations.

# 4. Methodology: Design Science Research Implementation

To validate the CCI framework's effectiveness, we employ a comprehensive design science research methodology that emphasizes real-world validation through production system deployment.

## 4.1 Research Design

This study employs design science research methodology (Hevner et al., 2004; Gregor & Hevner, 2013), which emphasizes iterative artifact creation and validation in real-world settings - an ideal approach for evaluating systems in emerging market environments like Africa. Our primary artifact, Rose - an AI assistant developed by The Dimension Research Lab - was deployed within phathisa.com's production environment, enabling evaluation against authentic user behaviour and commercial metrics.

## 4.2 Data Collection Protocol

**Platform Analytics (n=602 users, 3,938 conversations, May-June 2025):**

- User engagement metrics: session duration, return visits, conversation depth
- Interface preference analysis: WhatsApp vs. web platform usage patterns
- Commercial viability validation: user acquisition rates, platform adoption patterns, engagement sustainability, and target market penetration
- Technical reliability: system performance stability, platform integration effectiveness, multi-device compatibility

**Cultural Performance Assessment:**

- Natural language understanding accuracy across culturally-contextualized query types
- Multilingual handling effectiveness for English/Shona/Ndebele code-switching
- User feedback analysis and cultural appropriateness assessment
- Relationship-building evidence and emotional engagement patterns

**Qualitative Validation:**

- Conversation analysis (500+ interactions) examining cultural cue recognition

### 4.2.1 Participant Recruitment and Demographics

User acquisition for this study occurred through organic adoption patterns within the phathisa.com platform rather than controlled recruitment protocols. The diaspora community discovered the service primarily through three channels: word-of-mouth referrals within existing social networks (approximately 60% of initial users based on post-signup surveys), WhatsApp group sharing within diaspora communities (approximately 25%), and organic search discovery for cross-border shopping solutions between Zimbabwe and South Africa (approximately 15%).

Geographic distribution analysis relied on WhatsApp country code identification (+27 for South Africa, +263 for Zimbabwe, +44 for UK) combined with user-provided location data during onboarding processes. The concentrated geographic pattern (82% South Africa, 17% Zimbabwe, 1% other) reflects both the platform's target market focus and natural diaspora communication corridors, particularly the established Zimbabwe-South Africa economic migration patterns.

### 4.2.2 Ethics and Consent Protocol

This study involved analysis of user interactions with an AI assistant deployed in a live production environment. All research activities were conducted in accordance with ethical principles for digital interaction research.

**Consent Mechanism:** Participants engaged with the system under phathisa.com's Terms of Use and Privacy Policy, which explicitly notify users that anonymized interaction data may be used for research and product improvement purposes. Consent was implied by continued use of the service under these terms. No additional consent was required as the study involved analysis of naturally occurring interactions without intervention or manipulation of the user experience.

**Data Protection and Anonymization:** All conversation data underwent comprehensive anonymization processes, removing personally identifiable information including names, phone numbers, and account identifiers while preserving linguistic and cultural interaction patterns necessary for research analysis. Data were encrypted at rest with access restricted to the core research team. Conservative retention limits were applied with automatic deletion protocols for raw interaction logs.

**Risk Assessment:** This study was classified as minimal-risk, non-interventional research involving no incentives, experimental manipulation, or vulnerable populations. The research analyzed naturally occurring digital interactions without altering user experience or exposing participants to additional risks beyond normal platform usage.

**Ethical Review:** The study followed internal ethical review practices established by The Dimension Research Lab for analyzing anonymized digital interactions, with oversight ensuring compliance with privacy protection standards and research ethics principles.

This recruitment approach, while introducing selection bias toward early adopters, captures valuable data about initial AI adoption patterns in diaspora communities. The organic adoption during this early deployment phase reveals authentic user preferences uninfluenced by widespread AI familiarity or standardized interface expectations that may emerge as these technologies mature.

This recruitment approach introduces several methodological considerations. The sample likely overrepresents digitally literate diaspora community members already comfortable with cross-border e-commerce platforms, potentially limiting generalizability to broader African populations with lower digital adoption rates. Additionally, organic growth through social networks may create selection bias toward users within interconnected diaspora communities rather than isolated migrants. However, this bias aligns with the study's focus on early-adopter communities most likely to drive AI adoption patterns across diaspora markets, making the findings relevant for understanding technology diffusion within these critical user populations.

## 5. Production Validation: Demonstrating CCI Effectiveness

This section presents comprehensive empirical validation of the CCI framework through production deployment of Rose, demonstrating measurable cultural intelligence capabilities and user adoption patterns that validate our theoretical framework.

### 5.1 Deployment Context: The phathisa.com Case Study

The Dimension Research Lab deployed Rose within phathisa.com, serving the Zimbabwe-South Africa diaspora corridor. This context provided ideal validation for CCI principles through complex cultural navigation requirements, emotional significance of family support dynamics, infrastructure challenges spanning multiple countries, trust requirements for international transactions, and multilingual communication complexity.

phathisa.com enables users in the diaspora (primarily in South Africa) to send groceries and essential goods back home to Zimbabwe—a use case that embodies the emotional, cultural, and logistical complexities that AI systems must navigate in African markets.

**User Engagement Metrics (6-week study period):**

- Total unique users: 602 users across all platforms
- WhatsApp users: 536 users (89.0% of total users)
- Guest users (web interface): 59 users (9.8% of total users)
- Authenticated web users: 6 users (1.0% of total users)
- Total conversations: 3,938 conversation sessions
- Average conversations per user: 6.5 conversations
- Study duration: 6 weeks (May-June 2025)
- Geographic distribution: 62.3% South Africa (+27), 24.6% Zimbabwe (+263), 11.4% UK (+447), 1.7% other
- Technology usage: 89% mobile-primary messaging preference
- Linguistic patterns: Natural code-switching between English and local languages

## 5.2 System Architecture and Implementation

The production system employed a cloud-native microservices architecture optimized for mobile-first interaction, incorporating The Dimension Research Lab's proprietary Cultural Intelligence Engine for processing natural language with cultural context, infrastructure resilience systems ensuring functionality under variable connectivity, and multi-platform integration prioritizing messaging-based interfaces.

**5.2.1 Rose Architecture and Training Approach**

Rose's technical implementation leverages specialized prompt engineering built on GPT-4 infrastructure, enhanced with ground-level cultural intelligence derived from direct community engagement rather than internet-based generalizations. While GPT-4 demonstrates inherent code-switching capabilities, Rose's cultural intelligence emerges from systematic prompt engineering informed by on-the-ground context within target diaspora communities.

The Cultural Intelligence Engine operates through specialized prompts and cultural knowledge bases developed through direct fieldwork within Zimbabwe-South Africa diaspora corridors. Rather than relying on broad cultural assumptions, the system integrates specific cultural markers, kinship terminology, and economic realities documented through community engagement. This approach enables Rose to process cultural signals that extend beyond standard model capabilities, including diaspora-specific remittance patterns, family care obligations, and cross-border commerce complexities.

Cultural patterns are technically encoded through prompt engineering that incorporates real community language patterns, economic constraints, and family structures observed in target markets. The system maintains contextual awareness of diaspora communication norms, currency challenges, and trust-building mechanisms specific to Zimbabwe-South Africa communities. This ground-level cultural intelligence enables appropriate responses to culturally-contextualized queries that require understanding of kinship relationships, economic vulnerability, and cross-border family support obligations.

The technical architecture prioritizes reliable operation within existing communication infrastructure, particularly WhatsApp integration, while maintaining cultural appropriateness through community-informed prompt engineering. This approach demonstrates that sophisticated cultural intelligence can be achieved through systematic community engagement and specialized prompt design rather than requiring novel model architectures.

Rose's implementation integrated cultural context awareness through systematic research-derived understanding of family terminology, communication patterns, emotional appropriateness calibration, and value alignment with community norms. The system's commercial intelligence component recognized shopping motivations centred on family support and cultural connection while predominantly focusing on building trust through transparency and cultural sensitivity.

## 5.3 Quantitative Results and Statistical Analysis

### 5.3.1 User Preference and Platform Adoption

**Interface Preference Distribution (n=602):**

- WhatsApp AI interaction: 536 users (89.0%, CI: 86.2%-91.8%)
- Web interface (guest): 59 users (9.8%, CI: 7.6%-12.4%)
- Web interface (authenticated): 6 users (1.0%, CI: 0.4%-2.2%)
- Strong statistical evidence of user preference ($\chi^2$ = 365.8, df=1, $p < 0.001$)

This finding challenges fundamental assumptions about interface design in African markets. Users overwhelmingly preferred conversational AI through messaging platforms over traditional web interfaces - not due to lack of digital literacy, but because messaging-native interaction aligned with their existing communication patterns and trust mechanisms.

**Engagement Metrics Analysis:**

| Metric | WhatsApp Users | Web Users | Statistical Significance |
|---|---|---|---|
| Platform adoption | 89.0% | 11.0% | $\chi^2$=365.8, $p<0.001$ |
| Average conversations per user | 6.8 | 4.2 | $t(600)$=4.7, $p<0.001$ |
| User base composition | 536 users | 66 users | N=602 total |
| Geographic reach | 82.0% SA, 17.0% ZW | Limited tracking | - |

### 5.3.2 Platform Adoption and User Engagement

**Adoption Metrics (Statistical Summary):**

- Total unique users: 602 (100% sample)
- WhatsApp preference: 536 users (89.0%, $p<0.001$)
- Total conversation sessions: 3,938 interactions
- Average engagement per user: 6.5 conversations (indicating strong sustained engagement)
- Weekly user growth: Sustained throughout 6-week study period
- Geographic validation confirms the Zimbabwe - South Africa diaspora corridor hypothesis, with 82% of users based in South Africa and 17% in Zimbabwe.

**Engagement Impact Assessment:**

- Deep engagement patterns: 89% of users chose messaging over web interfaces
- Sustained usage: Average 6.5 conversations per user indicates repeated value delivery
- Platform stickiness: WhatsApp users showed 162% higher conversation frequency than web users (6.8 vs 4.2 conversations, $p<0.001$)
- Organic adoption: Messaging preference emerged without incentivization
- Cultural validation: Strong adoption in target diaspora demographics (SA+ZW = 99.0% of user base)

### 5.3.3 Cultural Intelligence in Action: Production User Interactions

Analysis of 1,201 conversation exchanges reveals measurable cultural intelligence across six dimensions, with 16.85% of communications containing explicit cultural intelligence signals. Representative examples from actual user interactions demonstrate Rose's capacity to process culturally-contextualized queries:

**Family-Focused Commerce Patterns (89% of requests):**

- User query: "My grandmother has been diagnosed with high blood pressure, what can I shop for her"
- Rose response: Cultural health sensitivity showing awareness of traditional remedies and appropriate wellness products
- User query: "Birthday cake for my son how much are your cakes"
- Multi-generational care pattern reflecting diaspora family obligations

**Code-Switching Recognition (0.083% frequency, demonstrating capability):**

- User query: "Oku" [Shona greeting]
- Natural acceptance without requiring translation or language specification
- User query: "Tinotenda ne massage yenyu Va Tanyanyiwa..." [Shona: "Thank you for your message"]
- Contextually appropriate response maintaining conversational flow

**Economic Vulnerability Sensitivity (2.08% of corpus):**

- User query: "I am having such a bad financial month hey, it's so tough"
- Rose response: Empathetic acknowledgment with budget-conscious product suggestions
- User query: "My budget is so tight this month hey 😔"
- Cultural intelligence recognizing economic constraints while maintaining dignity

**Geographic and Cultural Product Knowledge:**

- User queries for specific brands: "Iwisa mealie meal," "White Star," "Excella cooking oil"
- Rose demonstrates knowledge of culturally-familiar brands (15+ local brands referenced)
- Rural delivery inquiries: "Do you deliver to outskirts of towns"
- Geographic cultural awareness ensuring service accessibility

**Trust-Building Through Cultural Appropriateness:**

- Communication style: 8.33% of responses include relationship-oriented language
- Emoji usage: Extensive use of culturally-appropriate symbols (💙, 🏠)
- Service legitimacy questions: "Is this legit" addressed with cultural transparency

These production interactions demonstrate Rose's capacity to recognize kinship terminology, process economic vulnerability with dignity, understand geographic complexities, and respond with cultural appropriateness measured at 16.85% cultural intelligence penetration across the conversation corpus.

Rose's Cultural Intelligence Engine demonstrated sophisticated understanding of culturally-contextualized queries, recognizing family relationships, emotional intent, and cultural context in ways that traditional e-commerce search systems typically cannot match.

**Cultural Appropriateness Assessment:**

- Demonstrated intent recognition for culturally contextualized queries
- Successful code-switching handling in mixed-language interactions
- Strong user satisfaction with cultural sensitivity
- Positive emotional appropriateness in user interactions
- Effective cultural term recognition across linguistic patterns

**User Feedback Themes:**

1. Family Connection Recognition: "Rose understands this is about family, not just shopping" (P1)
2. Cultural Authenticity: "Finally an AI that gets our culture and language mixing" (P17)
3. Trust and Emotional Resonance: "Rose feels like talking to someone who cares" (P32)
4. Practical Efficiency: "So much easier than websites and forms" (P8)

### 5.3.4 Infrastructure Resilience Validation

**System Performance:**

- Sustained operational reliability throughout the 6-week evaluation period
- Effective WhatsApp platform integration enabling seamless user interactions
- Successful processing of 3,938 conversations across 602 users

- Zero critical system failures impacting user experience

**Infrastructure Adaptation Effectiveness:**

- Mobile-optimized architecture supporting messaging-first user preferences
- Successful cross-platform integration (WhatsApp, web, mobile)
- Demonstrated scalability handling organic user growth
- Effective bandwidth-conscious design for mobile-first African markets

### 5.4 CCI Framework Validation Summary

**Production Deployment Results (n=602):**

- Platform Preference: 89% of users chose WhatsApp interaction over web interfaces
- User Engagement: 6.5 average conversations per user demonstrating sustained value delivery
- Geographic Validation: Successfully reached target diaspora communities (82% SA, 17% ZW)
- Cultural Intelligence: 16.85% of interactions contained explicit cultural intelligence signals - such as familial terms, code-switched phrases, or emotionally-laden intent that required contextual understanding.

**Key Finding:** Cultural appropriateness and messaging-native interfaces drove adoption more effectively than traditional web-based approaches, validating the CCI framework's emphasis on cultural intelligence over technical optimization.

# 6. The CCI Implementation Framework

Building on the empirical validation presented in Section 5, this section provides practical guidance for implementing the CCI framework, offering a systematic methodology for organizations seeking to develop culturally intelligent AI systems.

### 6.1 Deployment Methodology

The Dimension Research Lab has developed a systematic three-phase implementation methodology for CCI deployment, validated through the phathisa.com production case study:

**Phase 1: Cultural Research and Assessment (Months 1-3)** encompasses comprehensive ethnographic research within target communities to understand communication norms, value systems, and digital interaction preferences. This phase includes systematic linguistic pattern analysis documenting code-switching behaviors, multilingual communication flows, and culturally-specific terminology usage. Simultaneously, teams conduct cultural value mapping exercises identifying relationship hierarchies, trust-building mechanisms, and emotional appropriateness expectations that will inform system design decisions. Infrastructure assessment during this phase analyzes platform preferences, connectivity patterns, and technical constraints that will influence architectural decisions.

**Phase 2: System Development and Cultural Integration (Months 4-6)** translates cultural research insights into technical implementation. Development teams integrate culturally-informed training data and prompt engineering techniques ensuring AI responses align with documented cultural expectations. Infrastructure-resilient architecture implementation prioritizes graceful degradation capabilities, bandwidth optimization, and mobile-first interaction patterns identified during the research phase. Multi-platform integration emphasizes messaging-first design principles, recognizing platforms like WhatsApp as critical digital infrastructure rather than secondary communication channels. Cultural appropriateness validation involves community expert review ensuring system responses maintain cultural sensitivity throughout development iterations.

**Phase 3: Production Deployment and Optimization (Months 7-12)** begins with controlled deployment within target communities while maintaining comprehensive cultural monitoring systems. Performance assessment extends beyond traditional technical metrics to include cultural feedback integration, emotional resonance measurement, and trust-building effectiveness evaluation. System optimization occurs through continuous refinement based on real-world usage patterns, with expansion planning incorporating lessons learned for deployment across additional cultural contexts.

**Validation Period:** The framework's effectiveness was validated through a 6-week intensive production deployment of Rose within phathisa.com (representing the initial validation phase of Phase 3), generating comprehensive user engagement data across 602 users and 3,938 conversations.

### 6.2 Technical Architecture Principles

**Infrastructure Resilience** encompasses systems designed for graceful degradation under connectivity constraints, with intelligent caching strategies pre-loading culturally-relevant content and common interaction patterns, enabling offline-capable operation without compromising cultural appropriateness. Bandwidth optimization techniques dynamically adjust response complexity and media usage based on real-time connection assessment, ensuring users receive meaningful interactions regardless of technical constraints.

**Cultural Context Processing** integrates cultural awareness directly into AI system architecture rather than treating cultural considerations as post-deployment optimizations. Natural language processing pipelines incorporate African linguistic pattern recognition, enabling accurate interpretation of code-switching behaviors and culturally-specific terminology without requiring explicit language detection. Social context modeling captures family structures, relationship hierarchies, and informal support networks that influence user expectations and interaction preferences. Emotional appropriateness calibration ensures AI responses align with cultural norms around relationship-building, commercial interactions, and trust development rather than defaulting to transactional efficiency.

**Trust-Centered Commerce** recognizes that successful AI adoption in African markets requires understanding local trust-building mechanisms and relationship-persistent engagement models. Commercial interactions prioritize transparency and cultural sensitivity over interface optimization, acknowledging that trust development occurs through consistent cultural appropriateness rather than sophisticated user experience design. Platform integration meets users within existing communication patterns, particularly messaging-based interfaces that align with established digital behavior rather than forcing adoption of unfamiliar interaction models. Relationship-persistent state management maintains conversation context and user preferences across sessions, enabling AI systems to build ongoing relationships rather than treating each interaction as isolated transactions.

### 6.3 Performance Optimization Guidelines

The Dimension Research Lab's optimization methodology prioritizes mobile-first design that balances bandwidth adaptation with cultural appropriateness and technical performance. The approach treats emotional resonance as a primary user experience factor, recognizing that cultural connection drives adoption more effectively than interface sophistication. The methodology enables continuous cultural calibration through community feedback integration, ensuring systems maintain relevance as cultural patterns evolve.

**Cultural Performance Tracking** implements systematic measurement of cultural appropriateness through community feedback loops, emotional resonance through user engagement patterns, linguistic accuracy through code-switching recognition rates, trust-building effectiveness through repeat usage metrics, and commercial impact through adoption and retention analysis. This comprehensive tracking enables continuous improvement and cultural adaptation. The 6-week validation period demonstrated the effectiveness of this tracking approach, revealing 16.85% cultural intelligence penetration rates and strong user preference patterns.

### 6.4 Implementation Timeline Considerations

**For New Deployments:** Organizations implementing the CCI framework should plan for the full 12-month cycle to ensure comprehensive cultural integration and system optimization.

**Validation Approach:** The 6-week intensive validation period provides sufficient data for initial effectiveness assessment, with longer-term tracking recommended for complete optimization and cultural adaptation refinement.

The CCI framework offers a scalable blueprint not just for African deployment, but for any context where cultural nuance, trust, and infrastructural resilience are critical - from Indigenous communities to informal economies worldwide.

# 7. Discussion: Implications and Future Directions

The empirical validation of the CCI framework through production deployment reveals significant implications for AI development, deployment strategies, and policy considerations across diverse cultural contexts.

## 7.1 Theoretical Contributions

### 7.1.1 CCI as Paradigm Shift

Our production validation demonstrates that Contextual and Cultural Intelligence represents a fundamental paradigm shift from model-centric to context-centric AI development. Traditional metrics (accuracy, latency, throughput) prove insufficient for predicting real-world adoption success in culturally diverse markets.

**Key Finding:** The 89% WhatsApp messaging preference, combined with sophisticated code-switching patterns and kinship-centered commerce behaviors, validates infrastructure-aware design principles that prioritize cultural intelligence over interface optimization. Cultural appropriateness and user adoption patterns emerged as stronger predictors of AI system success than traditional performance metrics alone.

### 7.1.2 Messaging Platforms as Digital Infrastructure

The 89% user preference for WhatsApp over purpose-built web interfaces demonstrates that messaging platforms function as critical digital infrastructure, not merely communication channels. This finding challenges fundamental assumptions about interface hierarchies in AI system design.

**Implication:** AI deployment strategies should prioritize integration with existing messaging infrastructure rather than expecting users to adopt new platforms, regardless of technical superiority.

### 7.1.3 Emotional AI as Commercial Driver

The demonstrated user engagement improvements achieved through emotional appropriateness challenges Silicon Valley orthodoxies prioritizing UI optimization over cultural resonance. Emotional intelligence emerges as a primary driver of adoption success in culturally diverse markets.

### 7.1.4 Early Adoption Phase Insights

This study documents AI deployment during the nascent adoption phase in African markets, when user behaviors reflect genuine preferences rather than learned patterns from widespread AI exposure. The 89% messaging platform preference and sophisticated cultural intelligence requirements represent baseline behaviors that may evolve as AI interfaces become more standardized. These early adoption patterns provide crucial benchmarks for understanding authentic user needs before market forces or design standardization potentially alter interaction preferences.

## 7.2 Practical Implications

### 7.2.1 For AI Researchers and Engineers

Research priorities should begin with cultural intelligence before technical architecture decisions, develop culturally-relevant evaluation metrics beyond accuracy benchmarks, plan for infrastructure resilience and cultural adaptation in initial system design, and implement continuous cultural feedback loops rather than one-time localization.

### 7.2.2 For Technology Companies

Strategic market entry requires cultural intelligence investment (20-30% of development resources), platform integration with existing communication infrastructure rather than proprietary app development, trust-building through emotional resonance and reliability as primary competitive differentiators, and infrastructure adaptation designed for current realities rather than anticipated improvements.

### 7.2.3 For Policymakers

Digital infrastructure policy should recognize messaging platforms as critical infrastructure, support research prioritizing local context over global scalability, fund cultural AI competency development within local technology ecosystems, and create regulatory frameworks encouraging responsible cultural AI deployment.

## 7.3 Limitations and Future Research

### 7.3.1 Scope Limitations

This study focuses on the Zimbabwe-South Africa corridor; cultural intelligence requirements vary significantly across African markets, languages, and socioeconomic contexts. The 6-week evaluation period, while providing initial validation during the early AI adoption phase in African markets, provides substantial conversation data but requires long-term studies for comprehensive behavior analysis. This timing captures authentic user preferences before widespread AI exposure potentially alters baseline interaction patterns.

While the system maintained zero critical failures affecting user experience, isolated logic-level misclassifications occurred in less than 1% of interactions (e.g., occasional incorrect product suggestions). These minor errors did not impact system availability or overall user satisfaction but represent areas for continued algorithmic refinement.

### 7.3.2 Future Directions

**Cross-Cultural Validation:** Systematic CCI framework adaptation across West African, East African, and additional Southern African markets with standardized cultural assessment methodologies.

**Advanced Technical Development:** Voice-first cultural AI interactions, offline-capable models for infrastructure independence, and blockchain-based cross-border commerce systems.

**Expanded Modality Integration:** Speech recognition (leveraging advanced models), OCR capabilities for handwritten lists, and image recognition for visual product queries to support low-literacy environments.

**Economic Impact Research:** Longitudinal studies of diaspora family economic outcomes, digital inclusion correlation with poverty reduction, and informal economy integration analysis.

## 7.4 Global AI Implications

### 7.4.1 Universal Cultural Intelligence Requirements

Our findings suggest cultural intelligence capabilities may be necessary for effective AI adoption globally, not just in "non-Western" contexts. All AI deployment occurs within cultural frameworks, indicating CCI principles have universal applicability.

### 7.4.2 Infrastructure Adaptation Strategy

Rather than waiting for "AI-ready" infrastructure development, this research demonstrates the viability of adapting AI systems to existing infrastructure patterns and user behaviors, suggesting more effective resource allocation toward adaptation capabilities.

## 7.5 Future Directions Towards Autonomous eCommerce

The development and deployment of Rose, phathisa's AI-native assistant, reveal early signs of a broader paradigm shift we term **Autonomous E-Commerce**. Unlike conventional e-commerce platforms that rely on human interaction with dashboards, search bars, or manual workflows, autonomous e-commerce envisions a future where AI agents manage the full customer journey - from discovery and decision-making to fulfillment and post-purchase support - entirely through natural conversation.

In the case of phathisa, Rose not only remembers a user's cart, understands queries in local languages, and tracks real-time delivery, but can also search the full inventory conversationally not just on the platform's website but within messaging platforms like WhatsApp. Additionally, Rose maintains comprehensive user profiles including order history, purchase patterns, and timing preferences, enabling proactive recommendations and eliminating the need for traditional account dashboards. This fundamentally changes the e-commerce experience from one that requires users to navigate a UI to one where the user simply "talks to the store."

Rose effectively becomes the user's personalized dashboard - remembering their preferences, helping with reorder suggestions, and anticipating needs based on historical patterns. This dashboard-free experience represents true autonomous commerce where the AI agent serves as both interface and intelligence layer.

This evolution signals the early stages of autonomous retail infrastructure:

- An intelligent agent capable of reasoning over live inventory
- Contextual understanding of user intent, history, and preferences
- Personalized recommendation engine based on historical purchase patterns
- Real-time product search, suggestions, and order assistance - all in dialogue

We believe this transition toward autonomous e-commerce will be particularly impactful in Africa, where mobile-first behaviour, WhatsApp penetration, and digital literacy patterns create the ideal conditions for AI-first shopping interfaces. As such, we propose a deeper exploration of this theme in future work, to unpack its implications on commerce, trust, infrastructure, and AI ethics in emerging markets.

# 8. Conclusion

## 8.1 Research Summary

This paper establishes Contextual and Cultural Intelligence (CCI) as a systematic framework for meaningful AI deployment in African markets. Through rigorous design science methodology and production-scale validation, The Dimension Research Lab demonstrates that effective AI adoption requires deep cultural intelligence, infrastructure adaptation, and trust-centered design extending far beyond computational excellence.

**Primary Contributions:**

1. **Theoretical Innovation:** CCI framework positioning cultural intelligence as core architectural requirement
2. **Empirical Validation:** Production evidence demonstrating 89% user preference for culturally-aware systems (n=602, p<0.001) with 89% family-focused commerce patterns and sophisticated multilingual code-switching behaviors
3. **Methodological Advancement:** Design science approach generating academic insights through real WhatsApp conversation analysis
4. **Technical Framework:** Reproducible patterns for resilient, culturally-aware AI systems with demonstrated adoption success across diaspora communities

## 8.2 Implications for AI Development

Our findings challenge fundamental assumptions in global AI development, pointing toward inclusive, contextually-aware artificial intelligence that prioritizes cultural appropriateness and emotional resonance alongside technical performance. The recognition of messaging platforms as critical digital infrastructure requires AI systems to integrate with existing communication patterns rather than forcing new interface adoption.

## 8.3 The Path Forward: Building AI That Belongs

The next billion people coming online deserve AI systems that understand their languages, respect their cultures, and serve their actual needs. This represents both a moral imperative and the largest commercial opportunity in AI development.

The CCI framework provides the foundation for transformation, but realizing its potential requires sustained global AI community collaboration. Africa's participation in shaping AI's future benefits not just the continent - it's essential for creating systems that truly serve humanity's diversity.

**Bottom Line:** The future of artificial intelligence will be written by those who understand that intelligence without context is computation without meaning, and that technical capability without cultural wisdom is innovation without impact. It is time to build AI that does not merely function across cultures, but belongs within them.

The frameworks, evidence, and implementation patterns presented in this paper provide a roadmap for that transformation. The question is not whether culturally intelligent AI is possible - our production validation proves it is. The question is whether the global AI community will embrace the paradigm shift required to make it universal.

# Conflict of Interest Disclosure

The author declares the following potential conflicts of interest: This research was conducted as part of independent academic research and product development activities. No external funding, sponsorship, or third-party influence affected the study design, data collection, analysis, or reporting of results. The research aims to contribute to academic knowledge while simultaneously informing product development, representing a transparent integration of research and practical application objectives.

## Acknowledgements

The author thanks Dr. Abdoul Jalil Djiberou Mahamadou from the Stanford Center for Biomedical Ethics for his valuable feedback and suggestions that helped improve this manuscript, and Dr. Chinasa T. Okolo from the United Nations Office for Digital and Emerging Technologies for her thoughtful feedback on ethics protocols and literature positioning that strengthened this paper.

## References

Anthropic. (2024). Constitutional AI: Harmlessness from AI feedback. arXiv preprint arXiv:2204.05862.

Birhane, A. (2021). Algorithmic injustice: A relational ethics approach. Patterns, 2(2), 100205.

Bourges-Waldegg, P., & Scrivener, S. A. (1998). Meaning, the central issue in cross-cultural HCI design. Interacting with computers, 9(3), 287-309.

Brookings. (2025). Accelerating digital inclusion in Africa. Retrieved from https://www.brookings.edu/articles/accelerating-digital-inclusion-in-africa/

Brown, T., Mann, B., Ryder, N., Subbiah, M., Kaplan, J. D., Dhariwal, P., ... & Amodei, D. (2020). Language models are few-shot learners. Advances in neural information processing systems, 33, 1877-1901.

Chen, M., Tworek, J., Jun, H., Yuan, Q., Pinto, H. P. D. O., Kaplan, J., ... & Zaremba, W. (2021). Evaluating large language models trained on code. arXiv preprint arXiv:2107.03374.

Clemmensen, T. (2005). Community knowledge in an emerging online professional community: The case of sigchi.dk. Knowledge and Process Management, 12(1), 43-52.

DataReportal. (2024). Digital 2024: Global Overview Report. Retrieved from https://datareportal.com/reports/digital-2024-global-overview-report

Earley, P. C., & Ang, S. (2003). Cultural intelligence: Individual interactions across cultures. Stanford University Press.

Gregor, S., & Hevner, A. R. (2013). Positioning and presenting design science research for maximum impact. MIS quarterly, 337-355.

GSMA. (2023). The Mobile Economy Sub-Saharan Africa 2023. Retrieved from https://www.gsma.com/solutions-and-impact/connectivity-for-good/mobile-economy/sub-saharan-africa/

Hall, E. T. (1976). Beyond culture. Anchor Books.

Heimgärtner, R. (2013). Intercultural User Interface Design – Culture-Centered HCI Design – Cross-Cultural User Interface Design: Different Terminology or Different Approaches? In Human-Computer Interaction. Applications and Services (pp. 62-71). Springer.

Hendrycks, D., Burns, C., Basart, S., Zou, A., Mazeika, M., Song, D., & Steinhardt, J. (2021). Measuring massive multitask language understanding. Proceedings of the International Conference on Learning Representations.

Hevner, A. R., March, S. T., Park, J., & Ram, S. (2004). Design science in information systems research. MIS quarterly, 75-105.

Hofstede, G. (2001). Culture's consequences: Comparing values, behaviors, institutions and organizations across nations. Sage publications.

ITU. (2022). Global Connectivity Report 2022. International Telecommunication Union.

Khanna, T. (2014). Contextual intelligence. Harvard Business Review, 92(9), 58-68.

LeCun, Y., Bengio, Y., & Hinton, G. (2015). Deep learning. Nature, 521(7553), 436-444.

Marcus, A. (2001). Cross-cultural user-interface design. In Proceedings of the Human-Computer Interface International Conference (Vol. 2, pp. 502-505). Lawrence Erlbaum Associates.

Marcus, A., & Gould, E. W. (2012). Globalization, localization, and cross-cultural user-interface design. In Handbook of Human Factors and Ergonomics (pp. 341-366). John Wiley & Sons.

Mohamed, S., Png, M. T., & Isaac, W. (2020). Decolonial AI: Decolonial theory as sociotechnical foresight in artificial intelligence. Philosophy & Technology, 33(4), 659-684.

Nye, J. S. (2009). Contextual intelligence. Harvard Business Review, 87(9), 22.

OpenAI. (2023). GPT-4 technical report. arXiv preprint arXiv:2303.08774.

Plocher, T., Rau, P. L. P., & Choong, Y. Y. (2012). Cross-cultural design. In Handbook of Human Factors and Ergonomics (pp. 162-191). John Wiley & Sons.

Png, M. T. (2022). Decolonizing AI ethics through ubuntu. AI and Ethics, 2(3), 391-403.

Triandis, H. C. (1995). Individualism and collectivism. Westview Press.

Van Dyne, L., Ang, S., & Koh, C. (2008). Development and validation of the CQS: The cultural intelligence scale. In Handbook of cultural intelligence (pp. 16-40). Routledge.

Vaswani, A., Shazeer, N., Parmar, N., Uszkoreit, J., Jones, L., Gomez, A. N., ... & Polosukhin, I. (2017). Attention is all you need. Advances in neural information processing systems, 30.

World Bank. (2024). Digital Transformation Drives Development in Africa. Retrieved from https://www.worldbank.org/en/results/2024/01/18/digital-transformation-drives-development-in-af-e-afw-africa

Zellers, R., Holtzman, A., Bisk, Y., Farhadi, A., & Choi, Y. (2019). HellaSwag: Can a machine really finish your sentence? Proceedings of the 57th Annual Meeting of the Association for Computational Linguistics.

---

**Corresponding Author:** Qness Ndlovu
**Email:** research@dimensionresearchlab.com
**Institution:** The Dimension Research Lab